# Spontaneous symmetry breaking of rapidly rotating stars in general relativity

Silvano Bonazzola, Joachim Frieben and Eric Gourgoulhon

Département d'Astrophysique Relativiste et de Cosmologie, Unité Propre 176 du CNRS,
Observatoire de Paris, F-92195 Meudon Cedex, France
*e-mail: bona,frieben,gourgoulhon@obspm.fr*

## ABSTRACT

We investigate the effects of general relativity upon the non-axisymmetric "bar" mode secular instability of rapidly rotating stars, i.e. the relativistic and compressible analog of the transition from Maclaurin spheroids to Jacobi ellipsoids. Our method consists in perturbing a stationary axisymmetric configuration, constructed by a 2-D general relativistic numerical code, and taking into account only the dominant terms in the non-axisymmetric part of the 3-D relativistic equations. For a polytropic equation of state, we have determined, as a function of the degree of relativity, the critical adiabatic index $\gamma_{\rm crit}$ above which rapidly rotating stars can break their axial symmetry. A by-product of the present study is the confirmation of the Newtonian value $\gamma_{\rm crit} = 2.238$ obtained by James (1964). We have also considered neutron star models contructed upon twelve nuclear matter equations of state taken from the literature. We found that five equations of state from this sample allow the symmetry breaking for sufficiently high rotation velocities. For the others, the Keplerian velocity (mass-shedding from the equator) is reached before the axisymmetry is broken. Rotating neutron stars that break their axial symmetry can be an important source of gravitational waves for the LIGO/VIRGO interferometric detectors.

*Subject headings:* Stars: Rotation — Stars: Neutron — Relativity — Instabilities — Gravitational Waves — Dense matter

## 1. Introduction

Triaxial instabilities of rotating neutron stars can play an important role as emission mechanisms of gravitational waves in the frequency range of the forthcoming LIGO/VIRGO interferometric detectors (see e.g. Bonazzola & Marck 1994 and references therein). Indeed, a just born rotating neutron star can spontaneously break its axial symmetry if the ratio of the rotational kinetic energy $T$ to the absolute value of the gravitational potential energy, $|W|$, exceeds some critical value. Alternatively, an old neutron star in a close binary system, accreting matter from its companion, may be spun up until the ratio $T/|W|$ is high enough to allow for the symmetry breaking. A steady state regime is achieved when the total accreted angular momentum is evacuated via gravitational radiation (Wagoner 1984), this process being known as *forced gravitational emission*.

There are important differences between the two astrophysical mechanisms mentioned above. The first one is that the energy radiated by the newborn neutron star is supplied by its rotational kinetic energy, whereas for the accreting neutron star it is supplied by the energy of the accreted matter. The second important difference concerns the temperature of these objects: the newborn neutron star is much hotter than the accreting one. Consequently the viscosity differs dramatically in the two cases, which causes the instability generating the symmetry breaking to be different. Indeed, as we recall below, two kind of secular instabilities plague rapidly rotating stars (cf. Schutz 1987 for a review): (i) the *Chandrasekhar-Friedman-Schutz instability* (Chandrasekhar 1970, Friedman & Schutz 1978, Friedman 1978, hereafter *CFS instability*) driven by gravitational radiation reaction, and (ii) the viscosity driven instability (Roberts & Stewartson 1963). Broadly speaking, the first effect is thought to be operative in newborn neutron stars and the second one in accreting neutron stars.

To be more precise, let us recall some classical results from the theory of rotating Newtonian homogeneous bodies. It is well known that a self-gravitating incompressible fluid rotating rigidly at some moderate velocity takes the shape of an axisymmetric ellipsoid: the so-called *Maclaurin spheroid*. At the critical point $T/|W| = 0.1375$ in the Maclaurin sequence, two families of triaxial ellipsoids branch off: the *Jacobi ellipsoids* and the *Dedekind ellipsoids*. The former are triaxial ellipsoids rotating rigidly around their smallest axis in an inertial frame, whereas the latter have a fixed triaxial figure in an inertial frame, with some internal fluid circulation at constant vorticity [see Chandrasekhar (1969) or Tassoul (1978) for a review of these classical results]. The Maclaurin spheroids are dynamically unstable for $T/|W| \geq 0.2738$. Thus the Jacobi/Dedekind bifurcation point $T/|W| = 0.1375$ is dynamically stable. However, in presence of some dissipative mechanism such as viscosity or gravitational radiation (CFS instability) that breaks the circulation or

– 2 –

angular momentum conservation, the bifurcation point becomes secularly unstable to the $l = 2, m = 2$ "bar" mode. Note also that a non-dissipative mechanism such as a magnetic field with a component parallel to the rotation axis breaks the circulation conservation (Christodoulou et al. 1995a) and may generate a spontaneous symmetry breaking. If one takes into account only the viscosity, the growth of the bar mode leads to the deformation of the Maclaurin spheroid along a sequence of figures close to some Riemann S ellipsoids[1] and whose final state is a Jacobi ellipsoid (Press & Teukolsky 1973). At the opposite, if the gravitational radiation reaction is taken into account but not the viscosity, the Maclaurin spheroid evolves close to another Riemann S sequence towards a Dedekind ellipsoid (Miller 1974). Though these behaviors have been rigorously demonstrated for the first time by the authors mentioned above, both of them may be qualitatively understood rather easily (cf. also Christodoulou et al. 1995b for a discussion in terms of phase transition). For the first one, it should be noticed that a Jacobi ellipsoid has a lower mechanical energy $T + W$ than the Maclaurin ellipsoid which has the same angular momentum and (rest) mass. Since viscosity dissipates mechanical energy but preserves angular momentum (and of course rest mass), the viscosity driven evolution of a Maclaurin spheroid is towards the Jacobi ellipsoid of the same mass and angular momentum, when this latter exists, i.e. when $T/|W| \geq 0.1375$. Note that the final state, the Jacobi ellipsoid, is rigidly rotating so that the viscous dissipation stops at this point. As concerns the CFS instability, it should be noted that gravitational radiation does not conserve the angular momentum, but the fluid circulation around the equator of the star (see e.g. Sect. 4.2 of Lai et al. 1994), as well as around any closed contour on a plane parallel to the equator. Now, for a given circulation and (rest) mass, a Dedekind ellipsoid has a lower mechanical energy than a Maclaurin spheroid. So the gravitational radiation driven evolution of a Maclaurin spheroid is towards the Dedekind ellipsoid which has the same mass and equatorial circulation. The evolution stops at the Dedekind ellipsoid because such a body does not emit any gravitational radiation (since it is stationary in the inertial frame at rest with respect to the star centre, its mass quadrupole moment does not vary). When both viscosity and gravitational radiation reaction are taken into account, their effects tend to cancel each other and therefore to stabilize the star (Lindblom & Detweiler 1977). The most extreme case occurs when the strengths of the two dissipative forces are exactly the same: the Maclaurin spheroids are then secularly stable up to the dynamical instability point $T/|W| = 0.2738$. When one of the two mecanisms is negligle with respect of the other, the critical value of $T/|W|$ is simply slightly higher than 0.1375 (see e.g. Fig. 10 of Lai & Shapiro 1995). To be complete, let us mention that the CFS instability has been found to occur before the bifurcation point $T/|W| = 0.1375$ for the modes with the azimuthal number $m$ higher than 2. In fact, the higher $m$, the lower the angular velocity for which the instability occurs [cf. Schutz (1987) or Lindblom (1992) for details]. Under the effect of gravitational radiation reaction solely, all rotating stars are thus unstable. In practice, viscosity suppresses the CFS instability for $m > 5$.

---

[1]The *Riemann S* family is formed by homogeneous bodies whose fluid motion can be decomposed into a rigid rotation around a principal axis and a uniform circulation whose vorticity is parallel to the rotation vector. Maclaurin, Jacobi and Dedekind ellipsoids are all special cases of Riemann S ellipsoids [for more details, cf. Chap. 7 of Chandrasekhar (1969) or Sect. 5 of Lai et al. (1993)].

The classical results recalled above have been extended to compressible fluids, modelled by a polytropic equation of state (hereafter EOS) by a number of authors. First of all Jeans (1919, 1928) has shown that a bifurcation point towards triaxial configurations can exist only if the adiabatic index $\gamma$ is larger than $\gamma_{\rm crit} \simeq 2.2$. The interpretation is that the EOS must be stiff enough for the bifurcation point to occur at an angular velocity lower than the maximum angular velocity $\Omega_{\rm K}$ for which a stationary solution exists. $\Omega_{\rm K}$ is reached when the centrifugal force exactly balances the gravitational force at the equator of the star, and for this reason is called the *Keplerian velocity*; if the star was forced to rotate at $\Omega > \Omega_{\rm K}$, it would lose some matter from the equator. By numerical calculations, James (1964) has refined Jean's result to

$$\gamma_{\rm crit} = 2.238 \qquad ({\rm James~1964}). \tag{1}$$

As concerns the secular instability in the compressible case, Ipser & Managan (1985) have shown that the $m = 2$ Jacobi-like bifurcation point has the same location along uniformly rotating sequences as the $m = 2$ Dedekind-like point, as in the incompressible case. The mutual cancellation of the viscosity and CFS instability still exists in the compressible case. For $\gamma < \gamma_{\rm crit}$, the CFS instability still exists for modes $l = m \geq 3$ (Managan 1985, Imamura et al. 1985) but not the viscous instability: if the viscosity is important, its effect is always stabilizing by acting against the CFS instability. Lindblom (1995) has shown that for a star of $1.5\,M_\odot$ constructed upon a polytropic EOS with $\gamma = 2$, the CFS instability is suppressed at temperatures $T < 5 \times 10^6$ K by the shear viscosity and at $T > 10^{10}$ K by the bulk viscosity. These results have been confirmed by Yoshida & Eriguchi (1995) [see also Sect. 2 of Friedman & Ipser (1992) and references therein]. To be complete, let us note that if the neutron star interior is superfluid, the CFS instability is suppressed by the "mutual friction" of its components (Lindblom & Mendell 1995).

Regarding the gravitational wave signal from rotating neutron stars that undergo the above triaxial instabilities, Ipser & Managan (1984) have examined the case of polytropic stars with $\gamma = 2.66$ and $\gamma = 3$ which have bifurcated along a triaxial Jacobi-like sequence, under the effect of the viscosity driven instability. Wagoner (1984) has computed the gravitational signal from an accreting neutron star — modelled by nearly spherical homogeneous objects — for the five lowest modes of the CFS instability. Recently, Lai & Shapiro (1995) have determined the gravitational wave form from newborn neutron stars — modelled as self-similar ellipsoids — undergoing the bar mode $l = m = 2$ of the CFS instability. Besides these studies where the instability mechanism is explicited and its growth computed, numerous estimates of gravitational radiation from rapidly rotating neutron stars have been published in which the deviation from axisymmetry, measured by the ellipticity $\epsilon$ in the equatorial plane, is assumed to take a fixed value [cf. New et al. (1995) and references therein]. Note that in these latter works, the source of the asymmetry $\epsilon$ is not necessarily the rotation induced instability, but can be a non-axisymmetric crystallization of the neutron star crust or pressure and magnetic stress anisotropies.

In view of all the results recalled hereabove, two questions that naturally arise are

1. Are the realistic[2] EOS of nuclear matter stiff enough to allow for the bar mode $l = m = 2$ instability of rotating neutron stars ?

2. What are the effects of general relativity on the symmetry breaking ?

As regards question 1, it should be noted that realistic EOS are far from being polytropic; in particular they are generally much softer in the external layers than in the core of the star. This can have some significant influence on the value of the Keplerian velocity $\Omega_{\rm K}$ and therefore on the existence of the bifurcation point. In this respect, defining a "mean adiabatic index" to determine if some realistic EOS allows the symmetry breaking by direct comparison with the polytropic case ($\gamma_{\rm crit} \simeq 2.24$), as certain authors did (Chau et al. 1992), seems highly questionable.

As regards question 2, it should be noted that all the works cited hereabove have been performed within the Newtonian theory of gravitation. Some Post-Newtonian studies (at the 1-PN level) of the Maclaurin and Jacobi ellipsoids have been completed (Chandrasekhar 1967, 1971, Bardeen 1971, Tsirulev & Tsvetkov 1982a,b). Chandrasekhar (1967) and Tsirulev & Tsvetkov (1982b) have notably shown that the Jacobi/Dedekind bifurcation point along a Post-Newtonian Maclaurin sequence occurs, for the same density, at higher angular velocity than in the Newtonian case. As regards the compressible (polytropic) case, preliminary results on the Post-Newtonian effects on the CFS instability have been published by Cutler & Lindblom (1992) and Lindblom (1995). No fully relativistic study of the bifurcation point has been performed yet. Now it must be stressed that neutron stars are relativistic objects and a Post-Newtonian approximation is not sufficient to describe neutron stars whose mass is larger than 1 $M_\odot$. An attempt has been made by Friedman et al. (1986) to define the kinetical energy of a fully relativistic rotating star by $T = J\Omega/2$, where $J$ is the angular momentum — which is defined unambiguously for an axisymmetric spacetime — and the gravitational potential energy by $W = Mc^2 - T - M_\mathrm{p}c^2$ where $M$ is the total mass-energy and $M_\mathrm{p}$ contains the fluid rest mass and internal energy. Friedman et al. then formed the ratio $T/|W|$ and assumed that triaxial instabilities set in at approximatively the same value of $T/|W|$ as in the Newtonian case.

---

[2] as opposed to polytropic

The purpose of the present paper is to answer questions 1 and 2 for the bar mode instability, i.e. to study the analog of the Maclaurin $\to$ Jacobi bifurcation point in the compressible and fully general relativistic frame. A basic assumption of our study is that of *rigid motion*, a notion that is well defined also in the relativistic case. Therefore our work is applicable when some mechanism is efficient to rigidify the motion. As detailed below, this can be shear viscosity, but also some magnetic field. The method we use consists in perturbing a fully relativistic axisymmetric configuration with a $l = m = 2$ mode, and to follow, within a given approximation, the growth or the decay of the perturbation.

The content of the article is as follows. In Sect. 2. we present the approximations used to tackle the problem and the resulting equations to be solved. Sect. 3. is devoted to the numerical method and to the tests of the numerical code. Results for relativistic polytropes are presented in Sect. 4., whereas Sect. 5. contains the results for twelve nuclear matter EOS from the literature. The concluding remarks are given in Sect. 6..

## 2. Basic assumptions and equations to be solved

### 2.1. Rigid rotation

#### 2.1.1. Astrophysical justification

A neutron star can be forced to rotate rigidly if the viscosity is high enough to damp out any deviation from uniform rotation. In this case, the CFS instability ("Dedekind-like" mode) is locked and the instability proceeds via the "Jacobi-like" mode. The competition between the two modes is governed by the ratio of the strength of the gravitational radiation reaction to the strength of viscous force. For a Maclaurin spheroid of average radius $R$ and mass $M$, this ratio is (Lindblom & Detweiler 1977)

$$X = 4.41 \times 10^{-13} \left(\frac{\nu}{\mathrm{cm}^2\mathrm{s}^{-1}}\right) \left(\frac{R}{10\ \mathrm{km}}\right)^2 \left(\frac{M}{1.4\ M_\odot}\right)^{-3}, \qquad (2)$$

where $\nu$ is the value of the kinematic viscosity. The viscosity of nuclear matter can have various origins and its value depends crucially on the temperature. For temperatures higher than $\sim 10^9$ K, the neutrons are not superfluid and the neutron-neutron scattering gives rise to the shear viscosity $\nu_{\rm n-n} = 1.9 \times 10^3 \, (\rho/10^{15} \, {\rm g\,cm^{-3}})^{5/4} \, (T/10^9 \, {\rm K})^{-2} \, {\rm cm^2\,s^{-1}}$ [Cutler & Lindblom (1987) fit of Flowers & Itoh (1976) result]. When reported in equation (2), this value leads to a ratio $X$ greater than one (CFS instability suppressed) for $T \lesssim 5 \times 10^4$ K, which is well below the threshold of the neutron normal state (i.e. non-superfluid) regime. In the superfluid regime ($T \lesssim 10^9$ K), the main contribution to the shear viscosity comes from the electron-electron scattering: $\nu_{\rm e-e} = 6.0 \times 10^3 \, (\rho/10^{15} \, {\rm g\,cm^{-3}}) \, (T/10^9 \, {\rm K})^{-2} \, {\rm cm^2\,s^{-1}}$ (Cutler & Lindblom (1987) fit of Flowers & Itoh (1976) result). With such a value $X > 1$ for $T \lesssim 10^5$ K. Note however that the ratio $X$ as defined by equation (2) — valid for an incompressible Newtonian fluid — provides only a crude estimate of the temperature for which viscous effects dominates. For compressible fluids (Newtonian polytropes), Ipser & Lindblom (1991) and Lindblom (1995) have shown that the CFS instability is suppressed by the shear viscosity for $T \lesssim 10^6 - 10^7$ K.

For high temperatures ($\sim 10^{10}$ K, new born neutron stars), the bulk viscosity resulting from the slowness of weak interaction processes may play a role. If the proton fraction $x_{\rm p}$ is lower than a critical value [$x_{\rm p}^{\rm crit} = 1/9$ for a (n,p,e$^-$) model of nuclear matter], only modified URCA processes are allowed and the resulting bulk viscosity has been computed by Sawyer (1979)[3] *in the case where the matter is transparent to neutrinos*, which requires $T \lesssim 2 \times 10^9$ K. In this regime the modified URCA bulk viscosity does not suppress the CFS instability, since Lindblom (1995) has shown that it is effective only for $T \gtrsim 10^{10}$ K, assuming a matter transparent to neutrinos (the most favorable case from the viscosity point of view). However, things change dramatically if *direct* URCA processes are allowed, i.e. if the proton fraction is greater than $\sim 1/9$. Note that modern EOS give a proton fraction which satisfies this requirement at high density (Lattimer et al. 1991). The kinematic bulk viscosity from direct URCA processes is $\nu_{\rm B} = 6.0 \times 10^{11} \, (\rho/10^{15} \, {\rm g\,cm^{-3}}) \, (T/10^9 \, {\rm K})^4 \, {\rm cm^2\,s^{-1}}$ in the (neutrino) transparent case and $\nu_{\rm B} = 1.0 \times 10^7 \, (\rho/10^{15} \, {\rm g\,cm^{-3}})^{-2/3} \, {\rm cm^2\,s^{-1}}$ in the opaque case (Haensel & Schaeffer 1992, with a perturbation timescale taken to be one millisecond). When these values are inserted in equation (2), it is found that the viscous effects dominate the gravitational radiation reaction ones for $T \gtrsim 10^9$ K in the transparent regime and never dominate in the opaque regime.

In conclusion, the Dedekind-like mode is blocked at low temperature ($T \lesssim 10^6$ K) by the shear viscosity resulting from electron-electron scattering and possibly around $10^9$ K by the bulk viscosity resulting from direct URCA processes, provided that the proton fraction is above $\sim 10$ %. Since at low temperature the shear viscosity dominates, one may assume that the star is rigidly rotating during the development of the non-axisymmetric instability. This does not stand any longer around $10^9$ K, because, even if the bulk viscosity is efficient in blocking the CFS instability, it cannot suppress any differential rotation.

---

[3] a misprint in the value of the bulk viscosity given in this reference has been corrected by Lindblom (1995).

A new born neutron star is very hot ($T \sim 10^{10}$ K) and the above discussion shows that neither shear nor bulk viscosity is effective to prevent the Dedekind-like mode. But another mechanism could prevent it, namely the magnetic field. Indeed, as noticed by Bonazzola & Marck (1994), assuming infinite conductivity, the shear of the fluid from differential angular velocity generates some toroidal magnetic field in addition to the poloidal one. In this manner, the excess of kinetic energy contained in differential rotation with respect to rigid rotation can be efficiently converted into magnetic energy, thereby enforcing rigid rotation.

### 2.1.2. *Geometrical translation*

To a very good approximation the neutron star matter is described by a perfect fluid, for which the stress-energy tensor takes the form

$$\mathbf{T} = (e+p)\mathbf{u} \otimes \mathbf{u} + p\,\mathbf{g}\ , \tag{3}$$

where $\mathbf{u}$ is the fluid 4-velocity, $e$ the fluid proper energy density, $p$ the fluid pressure and $\mathbf{g}$ the spacetime metric tensor.

Before the symmetry breaking, the spacetime generated by the rotating star can be considered as *stationary* and *axisymmetric*, which means that there exist two Killing vector fields, $\mathbf{k}$ and $\mathbf{m}$, such that $\mathbf{k}$ is timelike (at least far from the star) and $\mathbf{m}$ is spacelike and its orbits are closed curves. Moreover, in the case of rigid rotation, the spacetime is *circular*, which means that the 2-spaces orthogonal to both $\mathbf{k}$ and $\mathbf{m}$ are integrable in global 2-surfaces (cf. Carter 1973). This latter property considerably simplifies the study of rotating stars because some global coordinates $(t, r, \theta, \varphi)$ may be chosen so that the metric tensor components exhibit only one non-diagonal term ($g_{t\varphi}$) which is not vanishing. $t$ and $\varphi$ are coordinates associated with respectively the Killing vectors $\mathbf{k}$ and $\mathbf{m}$: $\mathbf{k} = \partial/\partial t$ and $\mathbf{m} = \partial/\partial \varphi$. The remaining coordinates $(r, \theta)$ span the 2-surfaces orthogonal to both $\mathbf{k}$ and $\mathbf{m}$. In all numerical works on rotating stars to date (see e.g. Bonazzola et al. 1993, hereafter BGSM, for a review) *isotropic coordinates* are chosen, for which the two-dimensional line element differs from the flat space one by a conformal factor $A^2$. In these coordinates, the components of the metric tensor are given by

$$g_{\alpha\beta}\,dx^\alpha dx^\beta = -N^2\,dt^2 + B^2 r^2 \sin^2\theta (d\varphi - N^\varphi\,dt)^2 + A^2 \left[dr^2 + r^2\,d\theta^2\right]\ , \tag{4}$$

where the four functions $N$, $N^\varphi$, $A$ and $B$ depend on the coordinates $(r, \theta)$ only, the coordinates $(t, \varphi)$ being associated with the Killing vector fields.

When the axisymmetry of the star is broken, the stationarity of spacetime is also broken. In the Newtonian theory, there is no inertial frame in which a rotating tri-axial object appears stationary, i.e. does not depend upon the time. It can be stationary only in a co-rotating frame, which is not inertial, so that the stationarity is broken in this sense. In the general relativistic case, where the notion of a global inertial frame is in general meaningless, a rotating tri-axial system is not stationary for it radiates away gravitational waves. Even if a co-rotating frame could be defined, the body could not be in a steady state in this frame, because it looses energy and angular momentum via gravitational radiation.

However, at the very point of the symmetry breaking, no gravitational wave has been emitted yet. For sufficiently small deviations from axisymmetry, we may neglect the gravitational radiation. Then, for a rigid rotation, there exists one Killing vector field **l**, which is proportional to the fluid 4-velocity **u**, hence

$$\mathbf{u} = \lambda \mathbf{l} \, , \tag{5}$$

where $\lambda$ is a strictly positive scalar function. Equation (5) is the definition of a *rigid motion* according to Carter (1979): (i) there exists a Killing vector field; (ii) the fluid 4-velocity is parallel to this Killing vector. In the axisymmetric and stationary case, the rigid motion corresponds to the constant angular velocity $\Omega := u^\varphi/u^t$, the Killing vector entering equation (5) being then

$$\mathbf{l} = \mathbf{k} + \Omega \, \mathbf{m} \, , \tag{6}$$

where **k** and **m** are the two Killing vectors defined above. The constancy of $\Omega$ ensures that **l** is a Killing vector too. The proportionality constant $\lambda$ of equation (5) is nothing else than the component $u^t$ of the 4-velocity **u**, where $t$ is the coordinate associated with the Killing vector **k**. Note that the Killing vector **l** is generally timelike close to the star and spacelike far from it (beyond the "light-cylinder").

In the non-axisymmetric case, **k** and **m** can no longer be defined as Killing vectors. We make instead the assumption that there exist (i) a vector field **k** which is timelike at least far from the star, (ii) a vector field **m**, which commutes with **k**, is spacelike everywhere and whose field lines are closed curves, (iii) a constant $\Omega$ such that the vector **l** defined by the combination (6) is a Killing vector and (iv) the fluid 4-velocity **u** is parallel to **l**. These hypotheses are the geometric translation of the approximation of rigid rotation and negligible gravitational radiation. The commutativity of **k** and **m** ensures that a coordinate system $(t, r, \theta, \varphi)$ can be found such that $\mathbf{k} = \partial/\partial t$ and $\mathbf{m} = \partial/\partial \varphi$.

## 2.2. First integral of motion

The properties that (i) the matter is described as a single constituent perfect fluid and (ii) the fluid motion is rigid in the sense defined above, enable one to find very simply a first integral of the equations of energy-momentum conservation $\nabla \cdot \mathbf{T} = 0$. We follow the general procedure described by Carter (1979). The starting point is the equations of motion written in the *canonical form* (Lichnerowicz 1967, Carter 1979, 1989)

$$\mathbf{u} \cdot (\nabla \wedge \boldsymbol{\pi}) = 0 , \tag{7}$$

where $\boldsymbol{\pi} := \mu \mathbf{u}$, $\mu$ being the baryon chemical potential, and $\nabla \wedge$ denotes the exterior differentiation, $\boldsymbol{\pi}$ being considered as a 1-form. Equation (7) is easily derived from the energy-momentum conservation equation $\nabla \cdot \mathbf{T} = 0$, with the form (3) of $\mathbf{T}$, the baryon number conservation relation $\nabla \cdot (n\mathbf{u}) = 0$ ($n$ being the fluid proper baryon density), and the First Law of thermodynamics which gives rise to (i) the equality between $\mu$ and the enthalpy per baryon $(e + p)/n$ and (ii) the Gibbs-Duhem relation $\nabla p = n \nabla \mu$. Now one has

$$\mathcal{L}_{\mathbf{l}} \boldsymbol{\pi} = \mathbf{l} \cdot (\nabla \wedge \boldsymbol{\pi}) + \nabla(\mathbf{l} \cdot \boldsymbol{\pi}) = 0 , \tag{8}$$

where $\mathcal{L}_{\mathbf{l}} \boldsymbol{\pi}$ denotes the Lie derivative of the 1-form $\boldsymbol{\pi}$ with respect to the vector field $\mathbf{l}$. The first equality in (8) results from Cartan's formula for the Lie derivative of differential forms, whereas the second equality follows from the fact that $\mathbf{l}$ is a symmetry generator of spacetime. Inserting the rigid motion relation (5) in equation (8) leads to

$$\lambda^{-1} \mathbf{u} \cdot (\nabla \wedge \boldsymbol{\pi}) + \nabla(\lambda^{-1} \mathbf{u} \cdot \boldsymbol{\pi}) = 0 . \tag{9}$$

By the canonical equation of motion (7), the first term of this relation vanishes. Replacing $\boldsymbol{\pi}$ by $\mu \mathbf{u}$ in the second term gives

$$\nabla(\lambda^{-1} \mu) = 0 , \tag{10}$$

from which it is immediatly deduced that

$$\lambda^{-1} \mu = \text{const.} \tag{11}$$

The scalar $\lambda$ can be expressed in terms of a certain Lorentz factor and a lapse function. Indeed let $\mathbf{n}$ be the unit future directed vector normal to the $t = \text{const.}$ hypersurfaces. By definition of the *lapse function* $N$ and the *shift vector* $\mathbf{N}$, $\mathbf{k} = N\mathbf{n} - \mathbf{N}$. The relations (5) and (6) then leads to

$$\mathbf{n} \cdot \mathbf{u} = \lambda N \mathbf{n} \cdot \mathbf{n} - \lambda \mathbf{n} \cdot \mathbf{N} + \lambda \Omega \mathbf{n} \cdot \mathbf{m} . \tag{12}$$

Now $\mathbf{n} \cdot \mathbf{N} = \mathbf{n} \cdot \mathbf{m} = 0$ since both $\mathbf{N}$ and $\mathbf{m}$ are tangent to the $t = $ const. hypersurfaces. The Lorentz factor between the observer of 4-velocity $\mathbf{n}$ (Eulerian observer) and the fluid comoving observer is

$$\Gamma = -\mathbf{n} \cdot \mathbf{u} \ . \tag{13}$$

Hence equation (12) gives

$$\lambda = \frac{\Gamma}{N} \ . \tag{14}$$

The first integral of motion (11) then becomes

$$\frac{\mu N}{\Gamma} = \mathrm{const.} \tag{15}$$

Let us introduce the logarithms

$$\nu := \ln N \tag{16}$$

$$H := \ln\left(\frac{\mu}{m_\mathrm{B} c^2}\right) , \tag{17}$$

where $m_\mathrm{B}$ is a mean baryon mass[4]. We call $H$ the log-enthalpy. For non-relativistic matter it reduces to the specific enthalpy disregarding the rest mass contribution. At the Newtonian limit, $\nu$ coincides with the gravitational potential. With these functions, the first integral of motion is written

$$H + \nu - \ln \Gamma = \mathrm{const.} \tag{18}$$

We note that this first integral has exactly the same functional form as in the axisymmetric stationary case [cf. Eq. (5.15) of BGSM]. The only difference is that now $H$, $\nu$ and $\Gamma$ are functions of $(r, \theta, \varphi - \Omega t)$ and not only of $(r, \theta)$.

### 2.3. Gravitational field equations

As discussed above, if gravitational radiation is neglected, we may consider that the star is rigidly rotating and that the spacetime exhibits some "helicoidal" symmetry (Killing vector field $\mathbf{l}$). Moreover, we are going to consider only the dominant term of the non-axisymmetric part of the gravitational potentials, which is the lapse function $N$, or its logarithm $\nu$. Indeed all the other gravitational potentials have matter source terms which are of the order of the fluid pressure or the product $\rho v/c$ (matter density × velocity) and we neglect the deviation from axisymmetry of these latter terms with respect to the deviation from axisymmetry of the matter density. In terms of a Post-Newtonian expansion, we consider the deviation from axisymmetry only at the 0-PN level. Accordingly, we keep the form (4) of the metric (with no other extra diagonal term than $g_{t\varphi}$), except that the metric coefficients are no longer functions of $(r, \theta)$ only, but become functions of $(r, \theta, \psi)$, where

$$\psi := \varphi - \Omega t \ . \tag{19}$$

---

[4]in this work, we use $m_\mathrm{B} = 1.66 \times 10^{-27}$ kg.

In the coordinates $(t, r, \theta, \psi)$, $t$ is the coordinate associated with the Killing vector **l**: **l** $= \partial/\partial t$ with $(r, \theta, \psi)$ held fixed. Within our approximation, the coordinate dependence of the metric coefficients is explicited as follows:

$$\begin{aligned} g_{\alpha\beta} \, dx^\alpha \, dx^\beta &= -N(r, \theta, \psi)^2 \, dt^2 + \frac{\tilde{B}(r, \theta)^2}{N(r, \theta, \psi)^2} \, r^2 \sin^2\theta \, [d\varphi - N^\varphi(r, \theta) \, dt]^2 \\ &\quad + \frac{\tilde{A}(r, \theta)^2}{N(r, \theta, \psi)^2} \left[ dr^2 + r^2 \, d\theta^2 \right] \ , \end{aligned} \tag{20}$$

where $\tilde{A}$ and $\tilde{B}$ are linked to the $A$ and $B$ factors entering equation (4) by $\tilde{A} = NA$ and $\tilde{B} = NB$. Note that the metric coefficients given by Eq. (20) are the components of the metric tensor with respect to the coordinates $(t, r, \theta, \varphi)$ and expressed as functions of the coordinates $(t, r, \theta, \psi)$. The gauge freedom arising as in any perturbative approach in general relativity is fixed by requiring the metric to take the form (20) in coordinates $(t, r, \theta, \varphi)$.

Note that in the weak gravitational field limit, expression (20) reduces to the well-known metric:

$$g_{\alpha\beta} \, dx^\alpha dx^\beta = -\left[1 + 2\nu(r, \theta, \psi)\right] \, dt^2 + \left[1 - 2\nu(r, \theta, \psi)\right] \left[dr^2 + r^2 \, d\theta^2 + r^2 \sin^2\theta \, d\varphi\right] \ , \tag{21}$$

from which the Einstein equations reduce to the Newtonian equations for the gravitational potential $\nu$.

The situation is summarized in Figure 1 by means of the Post-Newtonian parameter $\alpha$ which measures the intensity of the gravitational field: $\alpha = \max(\nu/c^2, v^2/c^2, p/m_B n c^2)$, and a parameter $\beta$ which measures the deviation from axisymmetry; $\beta$ is defined as the coefficient of the $\cos 2\psi$ term in the expansion of $\nu(r, \theta, \psi)$ at the stellar equator:

$$\nu\left(r = r_{\text{eq}}, \theta = \frac{\pi}{2}, \psi\right) = \text{const.} + \beta \cos 2\psi + \cdots \tag{22}$$

Alternatively, $\beta$ could have been chosen to be $(a_1 - a_2)/a_1$, where $a_1$ and $a_2$ are the stellar radii along the two principal axis of inertia in the plane perpendicular to the rotation axis. To summarize, our approach is an exact one (i.e. fully 3-dimensional, whatever the amplitude of the deviation from axisymmetry may be) at the Newtonian level. At the relativistic level, our formulation is exact when the deviation from axisymmetry is zero and takes into account only the 0-PN part of the non-axisymmetric terms.

In full generality (i.e. without any supposed spacetime symmetry nor any weak gravitational field assumption), an elliptic equation for the potential $\nu$ is obtained, via the 3+1 formalism of general relativity, by taking the trace of the projection of the Einstein equation onto the hypersurfaces $\Sigma_t$ defined by $t = $ const and making use of the Hamiltonian constraint equation. The result is [cf. e.g. Eq. (98) of York (1979) or Eq. (2.27) of Smarr & York (1978)]

$$\nu^{|i}_{\ |i} = 4\pi(E + S_i^{\ i}) - \nu^{|i}\nu_{|i} + K_{ij}K^{ij} - \frac{1}{N}\left(\frac{\partial K}{\partial t} + N^i K_{|i}\right) , \qquad (23)$$

where the latin indices $i, j$ run from 1 to 3, $\nu_{|i}$ denotes the covariant derivative of $\nu$ with respect to the 3-metric induced by $\mathbf{g}$ in the hypersurfaces $\Sigma_t$, $E$ and $S_{ij}$ are the fluid energy density and stress as measured by the Eulerian observer of 4-velocity $\mathbf{n}$ introduced in Sect. 2.2., $K_{ij}$ is the extrinsic curvature tensor of $\Sigma_t$, $K = K_i^{\ i}$, and the $N^i$'s are the components of the shift vector (cf. Sect. 2.2.). Within our approximation, equation (23) becomes

$$\Delta_3 \nu = 4\pi\frac{\tilde{A}^2}{N^2}(E + S_i^{\ i}) + \frac{\tilde{B}^2}{2N^4}r^2 \sin^2\theta \left[\left(\frac{\partial N^\varphi}{\partial r}\right)^2 + \frac{1}{r^2}\left(\frac{\partial N^\varphi}{\partial \theta}\right)^2\right]$$
$$- \frac{\partial \nu}{\partial r}\frac{\partial \ln \tilde{B}}{\partial r} - \frac{1}{r^2}\frac{\partial \nu}{\partial \theta}\frac{\partial \ln \tilde{B}}{\partial \theta} , \qquad (24)$$

where $\Delta_3$ stands for the Laplacian operator in 3-dimensional flat space:

$$\Delta_3 := \frac{\partial^2}{\partial r^2} + \frac{2}{r}\frac{\partial}{\partial r} + \frac{1}{r^2}\frac{\partial^2}{\partial \theta^2} + \frac{1}{r^2 \tan\theta}\frac{\partial}{\partial \theta} + \frac{1}{r^2 \sin^2\theta}\frac{\partial^2}{\partial \psi^2} . \qquad (25)$$

For the remaining gravitational potentials $N^\varphi$, $\tilde{A}$ and $\tilde{B}$, the equations are the same as in the axisymmetric case at our approximation level. They are fully explicited in BGSM. Here we simply recall their forms:

$$\tilde{\Delta}_3\left(N^\varphi\, r \sin\theta\right) = \sigma_{N^\varphi} \qquad (26)$$
$$\Delta_2\left(\tilde{B}\, r \sin\theta\right) = \sigma_{\tilde{B}} \qquad (27)$$
$$\Delta_2(\ln \tilde{A}) = \sigma_{\tilde{A}} , \qquad (28)$$

with

$$\Delta_2 := \frac{\partial^2}{\partial r^2} + \frac{1}{r}\frac{\partial}{\partial r} + \frac{1}{r^2}\frac{\partial^2}{\partial \theta^2} \qquad (29)$$

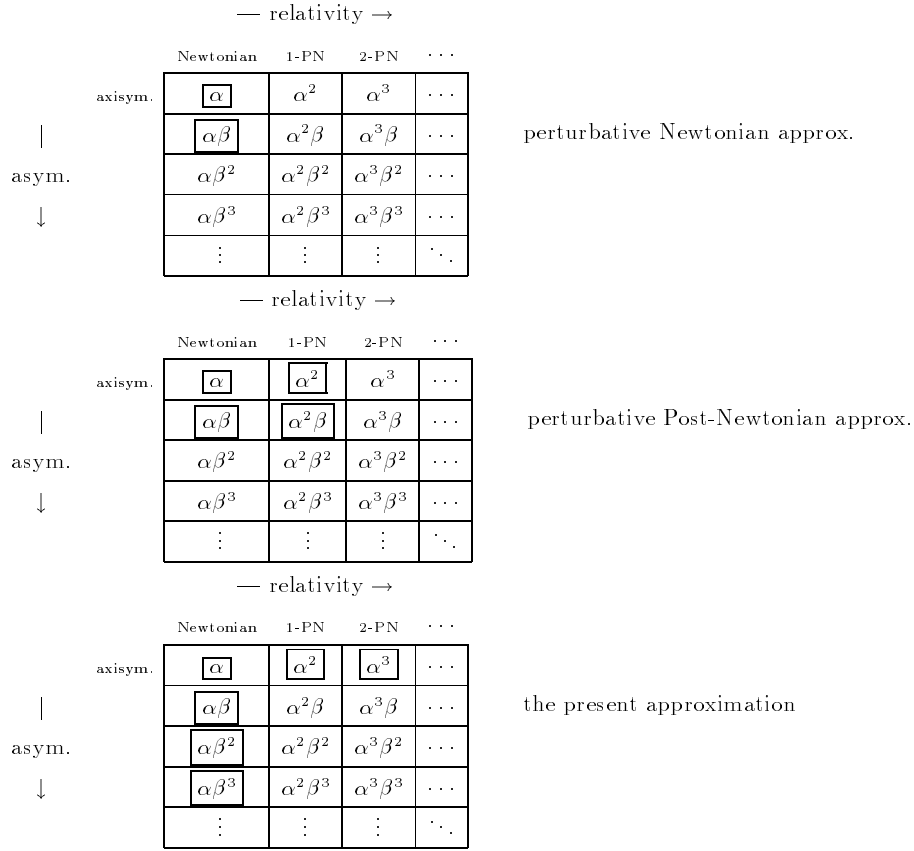

Fig. 1.— Various levels of approximation in terms of the Post-Newtonian parameter $\alpha \sim \nu/c^2$ and the asymmetry parameter $\beta$ (deviation from axisymmetry). "Perturbative" means that the deviations from axisymmetry are considered only at the linear order. The framed terms are those taken into account by each approach.

$$\tilde{\Delta}_3 := \frac{\partial^2}{\partial r^2} + \frac{2}{r}\frac{\partial}{\partial r} + \frac{1}{r^2}\frac{\partial^2}{\partial \theta^2} + \frac{1}{r^2 \tan\theta}\frac{\partial}{\partial \theta} - \frac{1}{r^2 \sin^2\theta} \ . \tag{30}$$

In the sources $\sigma$ of equations (26) – (28) appear some non-axisymmetric terms, such as the matter density or pressure, or the potential $\nu$. In order to be consistent with our approximation, the average of these terms over $\psi$ is used so that the solutions $N^\varphi$, $\tilde{A}$ and $\tilde{B}$ remain axisymmetric, as expressed in the line element (20).

As discussed in Sect. 3.1., the integration domain of the numerical code is the whole space. Therefore, the boundary conditions for the PDE system (24)-(28) must be imposed at spatial infinity, $r = +\infty$. In the case we consider (isolated star), the flat space values $\nu = 0$, $N^\varphi = 0$, $\tilde{B} = 1$ and $\tilde{A} = 1$ are imposed. In this way, it is guaranteed that the solution is asymptotically flat. One might fear that the 3-D solution of the system (24)-(28) is a spurious one in the sense of being made artificially 3-D by incoming gravitational radiation from infinity. This would not correspond to the astrophysical situation of an isolated rotating neutron star. Fortunately the solution of the system (24)-(28) with the boundary conditions listed above is free from any gravitational wave component, neither outcoming (this is physical since we are studying the very instant of the symmetry breaking) nor incoming. For instance, it can be seen very easily that the solution of Eq. (27) which satisfies the boundary condition $\tilde{B} = 1$ at $r = +\infty$ has the following asymptotic behavior $\tilde{B} \sim 1 + \text{const.} \times r^{-2} + O(r^{-3})$ (this comes from the fact that $\sigma_{\tilde{B}} = 0$ outside the star, cf. BGSM). If a gravitational wave was present, $\tilde{B}$ would necessarily contain a $r^{-1}$ part. In other words, the absence of any incoming gravitational wave is guaranteed by the fast decrease of the metric potentials at infinity.

### 2.4. Summary and adopted scheme for a solution

The equations displayed above are solved by an iterative scheme, starting from very crude initial conditions: a spherically symmetric star with constant density and a flat metric tensor. Given the density and pressure profiles, the Poisson-type gravitational field equations (24), (26), (27) and (28) are solved to get new values of $\nu$, $N^\varphi$, $\tilde{A}$ and $\tilde{B}$. The obtained $\nu$ is then put into the first integral of motion (18) to yield a new log-enthalpy $H$. From this latter, new density and pressure profiles are computed via the equation of state and a new iteration may begin[5]. For the first steps of this procedure, $\Omega$ is set to zero so that the solution remains spherically symmetric. At a certain step, the tenth say, $\Omega$ is switched on; consequently the solution becomes axisymmetric. As the number of iterations increases

---

[5]This kind of method is generally called a *Self-Consistent Field method*; cf. Sect. 5.5 of Tassoul (1978)

it converges to a certain state, which represents a stationary axisymmetric solution of Einstein equations. Then, at a given step, $J_0$ say, the following perturbative term is added to the potential $\nu$:

$$\delta\nu = -\varepsilon H_c \left(r \sin\theta \cos\psi\right)^2 \ , \tag{31}$$

where $H_c$ is the log-enthalpy at the centre of the star and $\varepsilon$ is a small constant of the order $10^{-6}$. The form (31) of $\delta\nu$ excites the "bar" mode $l = 2, m = \pm 2$. Following the first integral of motion (18), the enthalpy $H$ gets modified by the value $-\delta\nu$ and becomes non-axisymmetric, so the density and the pressure, via the EOS. At the following steps $J \geq J_0 + 1$ the perturbation (31) is switched off but the solution remains non-axisymmetric because of the non-axisymmetric parts of the matter density and pressure, as well as $N$, which appear in the source of the 3-D Poisson equation (24) for the potential $\nu$. The deviation from axisymmetry is measured by the parameter $\beta$ introduced in Sect. 2.3.. Then two different situations may happen:

1. $\beta$ tends to zero as the number of iterations increases, so that the final solution is axisymmetric.

2. $\beta$ grows as the number of iterations increases, so that the final solution deviates more and more from axisymmetry.

In case 1, we may state that the considered configuration is secularly stable for the bar mode when the deviation from axisymmetry occurs at the 0-PN level. Of course, it may be unstable to this mode thanks to higher order relativistic effects, which are not taken into account in the present study. So we cannot conclude about the secular stability of the configuration in case 1. In case 2, we conclude that the configuration is secularly unstable to the bar mode.

## 3. The numerical code and its tests

### 3.1. Numerical method

The iterative procedure described above has been implemented in a numerical code based on a spectral method. This code is a 3-dimensional extension of the code presented in BGSM and subsequently used to compute models of rapidly rotating neutron stars with various modern EOS of dense matter (Salgado et al. 1994a,b), as well as neutron star models with strong magnetic fields (Bocquet et al. 1995). We report to BGSM for technical details about the numerical code. Let us simply mention that our spectral method is based on an expansion of the scalar functions $f(r, \theta, \psi)$ in Chebyshev polynomials in $r$, in Legendre polynomials in $\theta$ and in Fourier series in $\psi$. Two spatial grids are used: one covering the star and the other one the space external to the star. This latter grid extends up to infinity thanks to the compactification induced by the change of variable $u = 1/r$.

### 3.2. Tests of the axisymmetric part

We have presented various tests passed by the code in the fully relativistic stationary axisymmetric case in BGSM, as well as in Bonazzola & Gourgoulhon (1994b). Some of these tests are provided by the virial identities GRV2 (Bonazzola 1973, Bonazzola & Gourgoulhon 1994a) and GRV3 (Gourgoulhon & Bonazzola 1994), the latter being a relativistic generalization of the Newtonian virial theorem. GRV2 leads to the $|1 - \lambda|$ error indicator introduced in BGSM. These virial identities allow to test each computation and not only the simplified ones for which an analytical solution is available. They notably control the convergence of the iterative procedure used in the code. In our calculations, the final value of the GRV2 or GRV3 error indicator are of the order of $10^{-14}$ in the spherically symmetric case and $10^{-6}$ (polytropic EOS) or a few $10^{-4}$ (tabulated EOS) in the rapidly rotating axisymmetric case. Let us recall that tabulated EOS introduce important numerical errors because they do not stricly obey to the thermodynamical relations (cf. Sect. 4.2 of Salgado et al. 1994a).

### 3.3. Tests of the non-axisymmetric part

Due to the lack of studies in the relativistic regime, we have tested the non-axisymmetric behaviour of our code only in the Newtonian limit, by comparison with previous works.

As stated in the introduction, James (1964) found that, for rotating Newtonian polytropes, a bifurcation from the axisymmetric sequence to some triaxial sequence exists only if the adiabatic index is greater than $\gamma_{\rm crit} = 2.238$. Using our code at the Newtonian limit, we have examined this classical result. Figure 2 reveals a distinct influence of the grid parameters on the actual value of $\gamma_{\rm crit}$. In the rotating case the surface of the flattened star does no longer coincide with the spherical boundary of the inner grid. A quantity such as the matter density is a continuous but not a smooth function across the stellar surface. Consequently the spectral method we use (expansion of the functions upon a set of analytic functions) approximates such a nonanalytic function rather poorly. This effect gets increasingly prohibitive for increasing values of $\gamma$ since surface derivatives become divergent for $\gamma > 2$. In the limit $\gamma \to \infty$, hence corresponding to the case of an incompressible fluid, the function itself is discontinuous and the spectral approximation becomes very bad. This is similar to the deficiency of Fourier series (a special kind of spectral method) in approximating discontinuous functions (*Gibbs phenomenon*). One should stress that any scheme that uses an expansion in the angular variables in terms of spherical harmonics is subjected to this constraint. From Fig. 2 we conclude that too small a number of grid points induces an artificial secular instability, hence the global increase of $\gamma_{\rm crit}$ for large values of $N_\theta$, while the additional oscillatory behaviour appears to be delicately related to the position of the star surface with respect to the grid points. From the behaviour of the critical index for large $N_\theta$ (left side of Fig. 2) we estimate the asymptotic value for an infinite number of grid points to be

$$\gamma_{\rm crit} = 2.238 \pm 0.002 , \qquad (32)$$

which coincides with James' value.

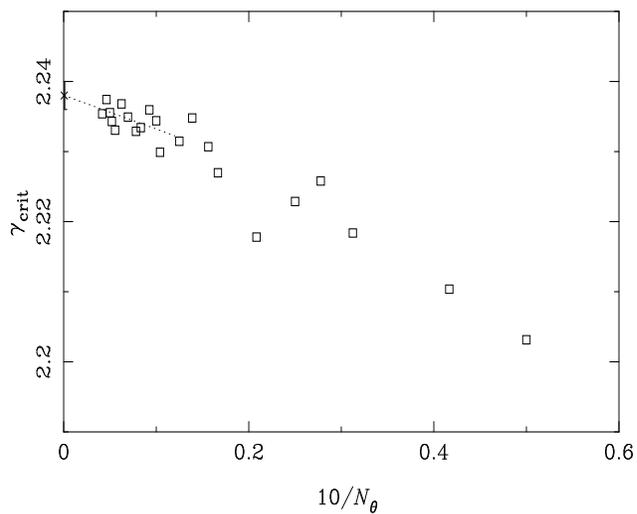

Fig. 2.— Variation of the critical adiabatic index $\gamma_{\rm crit}$ with the number of coefficients involved in the numerical method. $N_\theta$ is the number of Legendre polynomials; it ranges from 21 to 241, while the number $N_r$ of Chebyshev polynomials used to describe the radial part of the functions is set to $2N_\theta-1$.

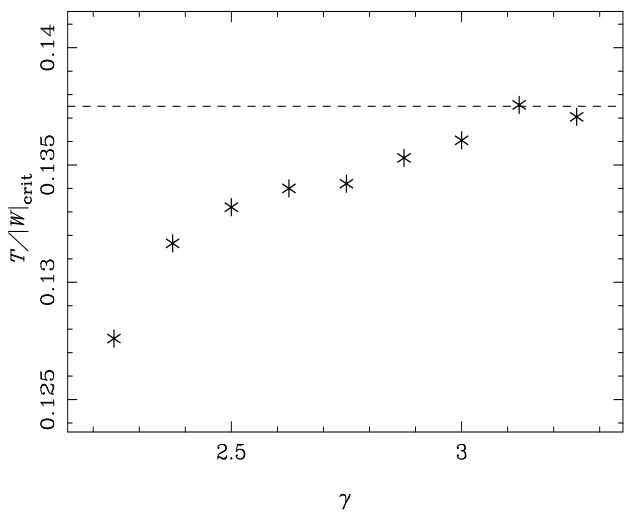

Fig. 3.— Ratio of the kinetic energy $T$ to the gravitational potential energy $W$ at the triaxial Jacobi-like bifurcation point along a sequence of rotating Newtonian polytropes, as a function of the adiabatic index $\gamma$. The dashed horizontal line corresponds to the theoretical value of $T/|W|$ for incompressible Maclaurin spheroids.

As we have explained above our code is not able to compute rapidly rotating *incompressible fluid* configurations, such as the Maclaurin and Jacobi ellipsoids, where the density profile is strongly discontinuous at the surface of the star, varying suddenly from its constant value to zero. For this reason we were not able to recover the classical results about the Maclaurin-Jacobi bifurcation point (cf. Sect. 1.). However, we searched the bifurcation point for polytropes of increasing abiabatic index $\gamma$, the (numerically unreachable) Maclaurin case corresponding to $\gamma = +\infty$. We computed the ratio $T/|W|_{\rm crit}$ of the rotational kinetic energy to the gravitational potential energy at the bifurcation point and we have checked that, as $\gamma$ increases, $T/|W|_{\rm crit}$ tends toward the classical value of the incompressible case: $T/|W|_{\rm crit}(\gamma = \infty) = 0.1375$ (Tassoul 1978). The results are depicted in Fig. 3, where $\gamma$ varies from $\gamma_{\rm crit} = 2.238$ to $\gamma = 3.25$ (for $\gamma$ greater than 3.25, the density profile steepens dramatically and the numerical error increases more and more).

Another test of the code consists in comparing results in the compressible polytropic case with previous numerical calculations. Ipser & Managan (1981) and Hachisu & Eriguchi (1982) have obtained numerical models of Newtonian triaxial rotating polytropes, analogous to the Jacobi ellipsoids. They did not determine $\gamma_{\rm crit}$[6] but performed calculations with fixed $\gamma \geq 2.66$. In table 1 we compare the location of the triaxial bifurcation point along a sequence of $\gamma = 3$ polytropes that we get with that obtained by the above authors. The agreement is better than 0.5% with the critical angular velocity of Hachisu & Eriguchi

---

[6] However, Ipser & Managan (1981) state that their results indicate that the critical adiabatic index lies somewhere in the range $2.22 \leq \gamma_{\rm crit} \leq 2.28$

(1982) and of the order of 2% with their critical value of $T/|W|$; with Ipser & Managan (1981), the agreement is better than 0.5% on both quantities.

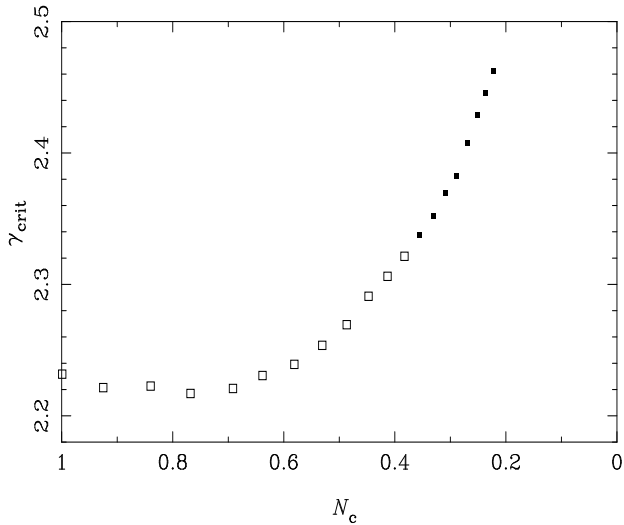

Fig. 4.— Critical polytropic index $\gamma_{\rm crit}$ as function of the lapse function $N_c$ measured at the centre of the star. Black boxes indicate configurations unstable with respect to radial oscillations.

## 4. Results for polytropes

The investigation of relativistic polytropic stars is interesting in several respects. First they represent a natural extension of former classical works restricted to the Newtonian case.

|  | $\dfrac{T}{|W|}$ | $\dfrac{\Omega^2}{4\pi G \rho_c}$ |
|---|---|---|
| this work | 0.1361 | 0.0676 |
| Ipser & Managan (1981) | 0.1364 | 0.0673 |
| Hachisu & Eriguchi (1982) | 0.1392 | 0.0679 |

Table 1: Comparison of the Jacobi-like bifurcation points along a sequence of Newtonian rotating polytropes with $\gamma = 3$. $\rho_c$ is the central density; other symbols are defined in the text.

Second, a polytropic EOS does not suffer from the thermodynamical inconsistency relative to tabulated EOS (cf. Sect. 4.2 of Salgado et al 1994a). It hence provides an approximate but consistent model for real stars which allows a first investigation of relativistic effects. Since Newtonian polytropes obey a scaling law, $\gamma_{\rm crit}$, the critical index for which the inset of secular triaxial instability coincides with the maximum rotating case, is a global constant. In the relativistic case however, relativistic effects are supposed to influence the symmetry breaking and the critical index will depend on an appropriate parameter measuring the relativistic character of the object. A well suited quantity is the value $N_c$ of the lapse function introduced in Sect. 2.2. at the centre of the star. We thus obtain a 2–dimensional parameter space for maximum rotating stars which is intersected by a curve representing the metastable configurations and thus separating the regions of stable and unstable stars respectively. Figure 4 shows the relativistic dependency of $\gamma_{\rm crit}$ ranging from the Newtonian to the extreme relativistic regime. In the moderately relativistic domain there appears a very slight decrease of $\gamma_{\rm crit}$ which can be seen more easily in Fig. 5 where additional points have been added. The negative slope reveals that the inset of relativistic effects tends to destabilize the star (within the limitations imposed by the approximate character of our theoretical approach), the maximum decrease of $\gamma_{\rm crit}$ being about 0.6 %. The discontinuities which appear in the course of $\gamma_{\rm crit}$ at this scale are due to the numerical effects explicited above for the Newtonian case and turned out to be closely correlated with the behaviour of the error estimator GRV2 (cf. Sect. 3.2.). They diminish for increasing $N_c$ thanks to the growing relativistic contributions to the source terms of the field equations which are better behaved in the spectral expansion. In the strong field region we observe a smooth growth of $\gamma_{\rm crit}$ beyond the maximum mass configurations due to the now persistently increasing stabilizing relativistic effects.

## 5. Results for realistic equations of state

We have determined in what condition the symmetry breaking may occur for rapidly rotating neutron stars built upon twelve EOS resulting from nuclear physics calculations. These "realistic" EOS are the same as whose used in Salgado et al (1994a) and we refer to this paper for a description of each EOS. The EOS are labeled by the following abreviations: PandN refers to the pure neutron EOS of Pandharipande (1971), BJI to model IH of Bethe & Johnson (1974), FP to the EOS of Friedman & Pandharipande (1981), HKP to the $n_0 = 0.17$ fm$^{-3}$ model of Haensel et al. (1981), DiazII to model II of Diaz Alonso (1985), Glend1, Glend2 and Glend3 to repsectively the case 1, 2, and 3 of Glendenning (1985), WFF1, WFF2 and WFF3 to respectively the AV$_{14}$ + UVII, UV$_{14}$ + UVII and UV$_{14}$ + TNI models of Wiringa et al. (1988), and WGW to the $\Lambda_{\rm Bonn}^{00}$ + HV model of Weber et al.

– 23 –

| EOS | $M_{\rm max}^{\rm stat}$ [$M_\odot$] | $M_{\rm max}^{\rm rot}$ [$M_\odot$] | $P_{\rm K}$ [ms] | $P_{\rm break}$ [ms] | $H_{\rm c,break}$ | $M_{\rm break}$ [$M_\odot$] |
|---|---|---|---|---|---|---|
| HKP | 2.827 | 3.432 | 0.737 | 1.215 | 0.161 | 1.80 |
| WFF2 | 2.187 | 2.586 | 0.505 | 0.755 | 0.30 | 1.951 |
| WFF1 | 2.123 | 2.528 | 0.476 | 0.728 | 0.27 | 1.736 |
| WGW | 1.967 | 2.358 | 0.676 | marginally stable | | |
| Glend3 | 1.964 | 2.308 | 0.710 | 0-PN stable | | |
| FP | 1.960 | 2.314 | 0.508 | 0.604 | 0.465 | 1.736 |
| DiazII | 1.928 | 2.256 | 0.673 | 0-PN stable | | |
| BJI | 1.850 | 2.146 | 0.589 | 0-PN stable | | |
| WFF3 | 1.836 | 2.172 | 0.550 | 0.714 | 0.325 | 1.909 |
| Glend1 | 1.803 | 2.125 | 0.726 | 0-PN stable | | |
| Glend2 | 1.777 | 2.087 | 0.758 | marginally stable | | |
| PandN | 1.657 | 1.928 | 0.489 | 0-PN stable | | |

Table 2: Neutron star properties according to various EOS: $M_{\rm max}^{\rm stat}$ is the maximum mass for static configurations, $M_{\rm max}^{\rm rot}$ is the maximum mass for rotating stationary configurations, $P_{\rm K}$ is the corresponding Keplerian period, $P_{\rm break}$ is the rotation period under which the symmetry breaking occurs, $H_{\rm c,break}$ is the central log-enthalpy at the bifurcation point and $M_{\rm break}$ is the corresponding gravitational mass. The EOS are ordered by decreasing values of $M_{\rm max}^{\rm stat}$.

(1991).

Our results are shown in table 2. For a given EOS, the axisymmetric rotating models form a two parameter family; each model can be labeled by its central energy density $e_c$ and its (constant) angular velocity $\Omega$. For a given value of $e_c$, $\Omega$ varies from zero to the Keplerian velocity $\Omega_K$. Following the method described in Sect. 2.4., we have searched for a symmetry breaking for configurations rotating at the Keplerian velocity. For five EOS (Glend1, Glend3, DiazII, BJI and PandN), no symmetry breaking was found, whatever the value of $\Omega_K$. These EOS are listed as "0-PN stable" in table 2. For two EOS (WGW and Glend2) the evolution of the parameter $\beta$ (cf. Sect. 2.4.) was not conclusive. These EOS are listed as "marginally stable" in table 2. A better numerical precision could lead to a definitive conclusion. For five EOS (HKP, FP, WFF1, WWF2 and WFF3), the bar mode reveals to be instable for some Keplerian velocities. Table 2 gives the period $P_{\text{break}}$, gravitational mass $M_{\text{break}}$ and central log-enthalpy $H_{c,\text{break}}$ (cf. Eq. (17)) of the configuration having the lowest angular velocity and for which the symmetry breaking occurs.

## 6. Conclusion

According to our study, the critical adiabatic index $\gamma_{\text{crit}}$ above which a rigidly rotating polytrope can undergo the bar mode instability varies with the relativistic character of the configuration, but not in a great range: from 2.238 for Newtonian bodies up to $\sim 2.4$ for very relativistic objects. As far as we know, the numerical result of James (1964) about the Newtonian value of $\gamma_{\text{crit}}$ has never been confirmed by any other published study, though it is more than thirty years old. The only references we are aware of on this subject are (i) the statement by Ipser & Managan (1981) that their results indicate that the critical adiabatic index lies somewhere in the range $2.22 \leq \gamma_{\text{crit}} \leq 2.28$ and (ii) a recent calculation by Skinner (1995, private communication) which confirms James' value. In this context, a by-product of the present work is the confirmation of James' result, namely $\gamma_{\text{crit}} = 2.238 \pm 0.002$.

As concerns realistic dense matter EOS, the stiffest ones allow the symmetry breaking, as expected, but the possibility for the symmetry breaking does not seem to be a monotonic function of the EOS stiffness. For instance the WFF3 EOS allow the symmetry breaking whereas it leads to a maximum mass of static neutron stars of only 1.84 $M_\odot$.

The tri-axial instability due to the Jacobi-like mode we have studied in this paper is likely to develop when some dissipation mechanism is efficient in enforcing the rigid rotation of the neutron star. This may occur in a very cold neutron star ($T \lesssim 10^6$ K, dissipation mechanism = shear viscosity). For a very hot neutron star ($T \sim 10^{10}$ K), the shear viscosity is inefficient and the tri-axial instability is triggered by the gravitational radiation reaction (CFS instability) and proceeds via a Dedekind-like mode, instead of the Jacobi-like one considered here. However, in this case, a strong internal magnetic field may rigidify the motion, blocking the Dedeking-like mode, but not the Jacobi-like one. A magnetic field gives also a stress tensor with "viscosity-like" off-diagonal terms and this could trigger an instability via the Jacobi-like mode. We plan to study the influence of the magnetic field on the stability of rotating neutron stars in a future paper. Hot and/or strongly magnetized neutron stars are expected to form at the end of a type II supernova collapse. Morever in this situation the neutron star rotation rate is likely to be increased by the accretion of high angular momentum material which has not been ejected by the supernova shock wave. If the axial symmetry is broken this could lead to a very strong pseudo-periodic gravitational wave signal in the frequency range of the forthcoming LIGO/VIRGO detectors. As the above results suggest, the positive detection of this signal would provide us with some constraints on the neutron star EOS.

We are grateful to Brandon Carter for fruitful discussions, and especially for his suggestion to depict our approximation by Fig. 1, to Pawel Haensel for providing us with some EOS tables, and to Dimitris M. Christodoulou and Jean-Pierre Lasota for reading the manuscript and making useful comments.
J. Frieben gratefully acknowledges financial support by the 'Gottlieb Daimler– und Karl Benz–Stiftung'.

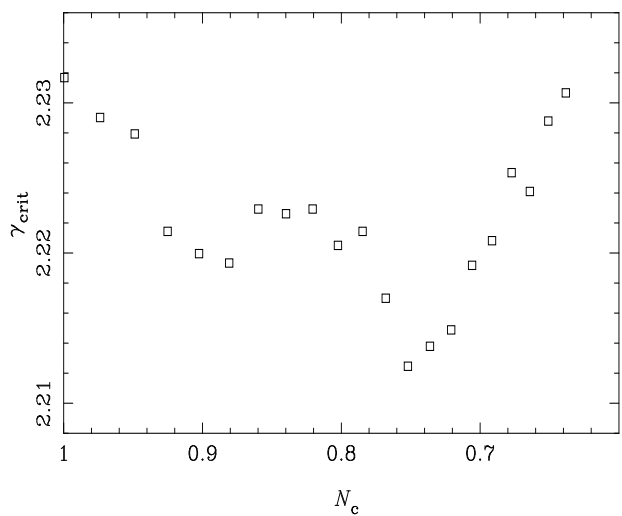

Fig. 5.— Same as Fig. 4 for the moderately relativistic region.